\documentclass[preprint,nofootinbib,showkeys]{revtex4-1}
\pdfoutput=1
\usepackage{amsmath}
\usepackage{cases}
\usepackage{amssymb}
\usepackage{graphicx,color}
\usepackage{hyperref}
\hypersetup{
	colorlinks = true,
	linkcolor = {blue},
	urlcolor = {black},
	citecolor = {blue}
}

\begin{document}

\title{Reheating neutron stars with the annihilation of self-interacting dark matter}

\author{Chian-Shu Chen$^{1,}$}
\email{chianshu@gmail.com}

\author{Yen-Hsun Lin$^{2,}$}
\email{yenhsun@phys.ncku.edu.tw}

\affiliation{$^1$Department of Physics, Tamkang University, New Taipei 251, Taiwan\\
	$^2$Department of Physics, National Cheng Kung University, Tainan 701, Taiwan}

\begin{abstract}
Compact stellar objects such as neutron stars (NS) are ideal places for capturing dark matter (DM) particles.
We study the effect of self-interacting DM (SIDM) captured by nearby NS that can reheat it to an appreciated surface temperature through absorbing the energy released due to DM annihilation.
When DM-nucleon cross section $\sigma_{\chi n}$ is small enough, DM self-interaction will take over the capture process and make the number of captured DM particles increased as well as the DM annihilation rate.
The corresponding NS surface temperature resulted from DM self-interaction is about hundreds of Kelvin and is potentially detectable by the future infrared telescopes.
Such observations could act as the complementary probe on DM properties to the current DM direct searches.

\end{abstract}

\maketitle

\section{Introduction}

Dark matter (DM) composes one-fourth of the Universe, however, its
essence is still elusive. Many terrestrial detectors 
are built to reveal the particle nature of DM either from measuring
the coupling strength between DM and the Standard Model (SM) particles \cite{Aad:2015zva,Abdallah:2015ter,Aalbers:2016jon,Akerib:2016vxi,Amole:2017dex,Akerib:2017kat,Aprile:2017iyp}
or the indirect
signal from DM annihilation in the space \cite{Aartsen:2014oha,Choi:2015ara,Aartsen:2016zhm,Aguilar:2015ctt,TheFermi-LAT:2017vmf,Ambrosi:2017wek}. But the definitive evidence
is yet to come.

A compact stellar object such as neutron star (NS) is a perfect
place to capture DM particles even when DM-nucleon cross section
$\sigma_{\chi n}$ is way smaller than the current direct search limits.
Investigations on DM in compact stellar objects are studied recently in Refs.~\cite{Kouvaris:2007ay,deLavallaz:2010wp,Kouvaris:2010vv,Kouvaris:2010jy,McDermott:2011jp,Guver:2012ba,Bramante:2013nma,Bramante:2017xlb,Tolos:2015qra,Leung:2011zz,Ellis:2017jgp,Ellis:2018bkr}.
Due to strong gravitational field, DM evaporation mass for NS is less than 10\,keV \cite{McDermott:2011jp}. Therefore, NS is sensitive to a broad spectrum of DM mass from 10\,keV to PeV, sometimes it can be even extended to higher mass region.
Unlike
the Sun, it loses its sensitivity to DM when $m_{\chi}\lesssim5\,{\rm GeV}$
as a consequence of evaporation \cite{Gould:1987ju,Chen:2014oaa}.
In the later discussion, we will
focus on the Weakly Interacting Massive Particle (WIMP) scenario with mass from MeV to hundreds of GeVs.

An old NS having age greater than billions of years could become a cold star after processing several cooling mechanism by emitting photons and neutrinos \cite{Shapiro:1983du,Page:2004fy,Potekhin:2014fla,Potekhin:2015qsa}.
However, if the residing DM particles in the NS can annihilate to
SM particles other than neutrinos, 
they will be absorbed by the host star and act like energy
injections to heat the star up
 \cite{Kouvaris:2007ay,deLavallaz:2010wp}.
In addition, recent literature also suggests
that the halo DM particles constantly bombard NS can deposit
their kinetic energy to the star. This is called dark kinetic heating \cite{Baryakhtar:2017dbj}.
These two contributions might prevent NS from inevitable cooling
as suggested by Refs.~\cite{Baryakhtar:2017dbj,Raj:2017wrv}.

Besides the DM-nucleon interaction, inconsistencies in the small-scale structure between
the observations and the \emph{N}-body simulations \cite{Navarro:1996gj,Moore:1994yx,Flores:1994gz,Randall:2007ph,Feng:2009hw,Walker:2011zu,Walker:2012td,BoylanKolchin:2011de,BoylanKolchin:2011dk,Elbert:2014bma,Bullock:2017xww}
imply the existence of self-interacting DM (SIDM)
\cite{Spergel:1999mh,Massey:2015dkw,Kahlhoefer:2015vua,Buckley:2009in,Aarssen:2012fx,Tulin:2012wi,Tulin:2017ara}.
The constraint given in Ref.~\cite{Kamada:2016euw,Robertson:2017mgj}
could mitigate such discrepancies as well as the diversity problem of the galactic rotation curves
\cite{Oman:2015xda,Elbert:2016dbb}. It brings us to
\begin{equation}
3\,{\rm cm}^{2}\,{\rm g}^{-1}\leq\sigma_{\chi\chi}/m_{\chi}\leq6\,{\rm cm}^{2}\,{\rm g}^{-1}\label{eq:xx_constraint}
\end{equation}
where $\sigma_{\chi\chi}$ is DM self-interaction cross section.
The resulting effect of DM self-capture in NS was considered insignificantly
due to it saturates quickly when the sum of the individual $\sigma_{\chi \chi}$ exceeds the geometrical area over which DM is thermally distributed \cite{McDermott:2011jp}. 
Its impact is unable to compete with the capture by DM-nucleon interaction when $\sigma_{\chi n}\gtrsim 10^{-50}\,{\rm cm^2}$.
However, current direct searches have put more stringent
limits on $\sigma_{\chi n}$ to test.
If it is small enough, DM self-capture will eventually take over \cite{Guver:2012ba}.
In this region, DM self-interaction will re-enhance the captured DM particles as well as the DM annihilation rate regardless how small $\sigma_{\chi n}$ is.
The corresponding energy injection increases consequently.
Hence, in the self-interaction dominant region, NS will experience a reheating effect with rising NS surface temperature.

An old, isolated NS nearby the Solar System with
	surface temperature with hundreds of Kelvin emits infrared that is a
very good candidate to pin down such reheating effects due to DM. 
The corresponding blackbody peak wavelength is potentially detectable
in the future telescopes, e.g.~the James Webb Space Telescope (JWST) \cite{Gardner:2006ky},
the Thirty Meter Telescope (TMT) \cite{Skidmore:2015lga} and
the European Extremely Large Telescope (E-ELT).
In the following context, we consider a nearby NS with age $t_{{\rm NS}}\simeq2\times10^{9}$
years, mass $M=1.44M_{\odot}$ where $M_{\odot}\approx1.9\times10^{33}\,{\rm g}$
is the Solar mass and radius $R=10.6\,{\rm km}$. It also has the halo density $\rho_{0}=0.3\,{\rm GeV}\,{\rm cm}^{-3}$,
the DM velocity dispersion $\bar{v}=270\,{\rm km}\,{\rm s}^{-1}$
and the NS velocity relative to the Galactic Center (GC) $v_{N}=220\,{\rm km}\,{\rm s}^{-1}$. For discussion convenience,
we will use natural unit $c=\hbar=k_{B}=1$ and $G=M_{P}^{-2}$ in this paper.


This paper is structured as follows: In Sec.~\ref{sec:NS_capture}
and Sec.~\ref{sec:cooling_heating}, we briefly review the
formalism of DM captured by NS and the cooling and heating
mechanism respectively. In Sec.~\ref{sec:numerical_result}, numerical results
are presented as well as the discussion on the reheating effect.
The implication of $T_{\rm sur}$ for DM direct searches is discussed too.
The work is summarized in Sec.~\ref{sec:summary}.

\section{DM captured by neutron star\label{sec:NS_capture}}

\subsection{DM evolution equation}\label{sec:DM_evo}

When the halo DM particles scatter with NS and lose significant
amount of energies, they will be gravitationally bounded in the star.
The evolution of DM number $N_\chi$ in NS can be characterized by the differential equation
\begin{equation}
\frac{dN_{\chi}}{dt}=C_{c}+C_{s}N_{\chi}-C_{a}N_{\chi}^{2}\label{eq:dN}
\end{equation}
where $C_{c}$ is the capture rate due
to DM-nucleon interaction, $C_{s}$ the DM self-capture rate due to DM self-interaction and $C_{a}$ the
DM annihilation rate. A general solution to $N_{\chi}$ is given by
\begin{equation}
N_{\chi}(t)=\frac{C_{c}\tanh(t/\tau)}{\tau^{-1}-C_{s}\tanh(t/\tau)/2}\label{eq:Nx}
\end{equation}
where $\tau=1/\sqrt{C_{c}C_{a}+C_{s}^{2}/4}$ is the equilibrium timescale.
In the case of $t\gg\tau$, $dN_{\chi}/dt=0$ where $N_\chi$ reaches the steady state. Hence we have
\begin{equation}
N_{\chi}(t\gg\tau)\equiv N_{\chi,{\rm eq}}=\sqrt{\frac{C_{c}}{C_{a}}}\left(\sqrt{\frac{R}{4}}+\sqrt{\frac{R}{4}+1}\right)\label{eq:Neq}
\end{equation}
where
\begin{equation}
R\equiv\frac{C_{s}^{2}}{C_{c}C_{a}}\begin{cases}
\gg1, & \textrm{\ensuremath{C_s}-dominant}\\
\ll1, & \textrm{\ensuremath{C_c}-dominant}
\end{cases}\label{eq:R}
\end{equation}
Thus, $R$ signifies how crucial that the DM self-capture
is in the DM evolution in NS.
Additionally, we can
obtain two solutions to $N_{\chi}$ when $dN_{\chi}/dt=0$, by examining Eq.~(\ref{eq:dN}),
\begin{equation}
N_{\chi,{\rm eq}}^{R\ll1}=\sqrt{\frac{C_{c}}{C_{a}}}\quad{\rm and}\quad N_{\chi,{\rm eq}}^{R\gg1}=\frac{C_{s}}{C_{a}}\label{eq:Neq2}.
\end{equation}
That means, either the capture is dominated by $C_{c}$ or $C_{s}$ that could accumulate the same amount of DM particles in NS.

\subsection{Rates of DM capture and annihilation}

The capture rate due to DM scattering with target neutrons in NS
is given by \cite{McDermott:2011jp}
\begin{equation}
C_{c}=\sqrt{\frac{6}{\pi}}\frac{\rho_{0}}{m_{\chi}}\frac{v_{{\rm esc}}(r)}{\bar{v}^{2}}\frac{\bar{v}}{1-2GM/R}\xi N_{n}\sigma_{\chi n}^{{\rm eff}}\left(1-\frac{1-e^{-B^{2}}}{B^{2}}\right)\label{eq:Cc}
\end{equation}
where $\rho_{0}$ is the DM density, $\bar{v}$ the DM velocity dispersion,
$N_{n}=M/m_{n}$ the total number of target neutrons in NS, and
$M$ and $R$ are the mass and radius of NS respectively.
The suppression factor $\xi=\delta p/p_{F}$ is due to the neutron
degeneracy effect. The momentum transfer in each scattering is $\delta p\simeq\sqrt{2}m_{r}v_{{\rm esc}}$ where $m_{r}=m_{\chi}m_{n}/(m_{\chi}+m_{n})$
the reduced mass and $v_{{\rm esc}}\simeq1.8\times10^{5}\,{\rm km}\,{\rm s}^{-1}$.
Since the DM-nucleon cross section $\sigma_{\chi n}$ cannot exceed the geometric limit that
is given by $N_{n}\sigma_{c}=\pi R^{2}$ where $\sigma_{c}\simeq2\times10^{-45}\,{\rm cm}^{2}$
is the critical cross section for DM-nucleon in NS. Thus, $\sigma_{\chi n}^{{\rm eff}}\equiv\min(\sigma_{\chi n},\sigma_{c})$ is the effective DM-nucleon
	cross section.
The last factor $B^{2}\equiv(3/2)(v_{{\rm esc}}^{2}/\bar{v}^{2})\beta_{-}$ where $\beta_{-}=4m_{\chi}m_{n}/(m_{\chi}-m_{n})^{2}$.
Unless $m_{\chi}\gtrsim10\,{\rm TeV}$, the term in the parentheses
is roughly unity.

Another way of capture is due to the halo DM particle scatters with
the trapped DM particle. This is DM self-capture and is given by \cite{Guver:2012ba,Zentner:2009is}
\begin{equation}
C_{s}=\sqrt{\frac{3}{2}}\frac{\rho_{0}}{m_{\chi}}v_{{\rm esc}}(R)\frac{v_{{\rm esc}}(R)}{\bar{v}}\langle\hat{\phi}_{\chi}\rangle\frac{{\rm erf}(\eta)}{\eta}\frac{1}{1-2GM/R}\sigma_{\chi\chi}\label{eq:Cs}
\end{equation}
where $v_{{\rm esc}}(R)$ is the escape velocity at the surface of
NS. For a rather conservative calculations, we take
$\langle\hat{\phi}_{\chi}\rangle=1$ \cite{Guver:2012ba}.
The quantity $\eta=\sqrt{3/2}(v_{N}/\bar{v})$ where $v_{N}=220\,{\rm km\,s^{-1}}$ is the
NS velocity relative to GC.

Usually the term $C_s N_\chi$ in Eq.~(\ref{eq:dN}) is proportional to $N_\chi \sigma_{\chi \chi}$,
	but it cannot grow arbitrarily as $N_\chi$ increases.
	The sum of individual $\sigma_{\chi \chi}$ never surpasses the geometric area over the DM particles are thermally distributed in the NS.
	The geometric area is characterized by the thermal radius $r_{\rm th}$ \cite{McDermott:2011jp}:
	\begin{equation}
	r_{{\rm th}}=\sqrt{\frac{9T_\chi}{4\pi G\rho_{n}m_{\chi}}}\approx24\,{\rm cm}\,\left(\frac{T_{\chi}}{10^{5}\,{\rm K}}\cdot\frac{100\,{\rm GeV}}{m_{\chi}}\right)^{1/2}\label{eq:Rth}
	\end{equation}
where $T_{\chi}$ is the DM temperature and such limitation is called the geometric limit for DM self-capture in NS.
Therefore, for any $\sigma_{\chi\chi}$ range given in Eq.~(\ref{eq:xx_constraint}), the term $N_{\chi}\sigma_{\chi\chi}$ must not larger than $\pi r_{{\rm th}}^{2}$.
Qualitative,
	we take Eq.~(\ref{eq:xx_constraint}) as our initial input for $\sigma_{\chi \chi}$
	in the numerical calculation.
However, it only serves the purpose of how fast $N_\chi \sigma_{\chi \chi}$
approaching  $\pi r_{\rm th}^2$.
After reaching this saturation, the initial $\sigma_{\chi \chi}$ input is irrelevant.
Though $N_\chi$ could grow thenceforth, the overall quantity $N_\chi \sigma_{\chi \chi}$ remains $\pi r_{\rm th}^2$. This procedure is carried out by our numerical program.
As a remark, such saturation for DM self-capture always happens.
If it is not considered in the calculation,
$N_\chi$ will be highly overestimated.

When more and more DM particles accumulate in the NS, the chance of DM annihilation becomes appreciated. Thus,
\begin{equation}
C_{a}\approx\frac{\langle\sigma v\rangle}{4\pi R^{3}/3}\label{eq:Ca}
\end{equation}
and the total annihilation rate is
\begin{equation}
\Gamma_{A}=\frac{1}{2}C_{a}N_{\chi}^{2}(t).\label{eq:GammaA}
\end{equation}
In the later discussion, we will use thermal relic annihilation cross
section $\langle\sigma v\rangle=3\times10^{-26}\,{\rm cm}^{3}\,{\rm s}^{-1}$.



\section{Neutron star cooling and energy injection due to DM annihilation\label{sec:cooling_heating}}

After the birth of NS, it undergoes the cooling mechanism due to
neutrino and photon emissions \cite{Shapiro:1983du,Page:2004fy}. Nonetheless, if the residing
DM particles in NS can annihilate, the annihilation products will be absorbed
and act like energy injections
to heat the host star up. The NS interior temperature $T_{\rm int}$ can be described by the following
differential equation
\begin{equation}
\frac{dT_{\rm int}}{dt}=\frac{-\epsilon_{\nu}-\epsilon_{\gamma}+\epsilon_{\chi}}{c_{V}}\label{eq:dT}
\end{equation}
where $\epsilon_{\nu,\gamma,\chi}$ are the emissivities due to neutrino emission,
photon emission and DM respectively. They are given
by \cite{Kouvaris:2007ay,Shapiro:1983du}
\begin{equation}
\epsilon_{\nu}\approx1.81\times10^{-27}\,{\rm GeV}^{4}\,{\rm yr}^{-1}\,\left(\frac{n}{n_{0}}\right)^{2/3}\left(\frac{T_{\rm int}}{10^{7}\,{\rm K}}\right)^{8}
\end{equation}
where 
$n\simeq 3.3\times10^{38}\,{\rm cm^{-3}}$ is the NS baryon number density
and $n_{0}\simeq 0.17\,{\rm fm^{-3}}$ the baryon density for the nuclear matter \cite{Kouvaris:2007ay}. It is therefore $n/n_0\simeq 2.3$.

Since the NS outer 
	envelope shields us from observing $T_{\rm int}$ directly. We can only observe the luminosity $L_\gamma$ emitted from this envelope.
	The corresponding temperature can be inferred from Stefan-Boltzmann's law that is defined as the NS surface temperature $T_{\rm sur}$ where $L_\gamma=4\pi R^2 \sigma_{\rm SB}T^4_{\rm sur}$. A relation that connects $T_{\rm int}$ and $T_{\rm sur}$
 is given by \cite{Page:2004fy,Gudmundsson1982,Gudmundsson1983}
\begin{equation}
T_{{\rm sur}}=0.87\times10^{6}\,{\rm K}\,\left(\frac{g_{s}}{10^{14}\,{\rm cm}\,{\rm s}^{-2}}\right)^{1/4}\left(\frac{T_{\rm int}}{10^{8}\,{\rm K}}\right)^{0.55}\label{eq:Tsur}
\end{equation}
where $g_{s}=GM/R^{2}=1.85\times10^{14}\,{\rm cm}\,{\rm s}^{-2}$
is the surface gravity.
In general, $T_{\rm sur}$ is
	lower than $T_{\rm int}$. However,
	when $T_{\rm int}\lesssim 3700\,{\rm K}$, the distinction between the two
	becomes negligible \cite{Page:2004fy}. Applying Eq.~(\ref{eq:Tsur}) to
obtain $T_{\rm sur}$ from $T_{\rm int}$ is unnecessary when $T_{\rm int}\lesssim 3700\,{\rm K}$.

On the other hand, $L_\gamma$ is also responsible for the energy
loss due to photon emission.
Hence we have the effective photon emissivity
\begin{subnumcases}
{\epsilon_{\gamma}=\frac{L_{\gamma}}{(4/3)\pi R^{3}}\approx}
{ 2.71\times10^{-17}\,{\rm GeV^{4}\,yr^{-1}}\left(\frac{T_{\rm int}}{10^{8}\,{\rm K}}\right)^{2.2} } & $ T_{\rm int}\gtrsim3700\,{\rm K} $, \label{eq:gtr_th}
\\
{ 2.56\times10^{-9}\,{\rm GeV^{4}\,yr^{-1}}\left(\frac{T_{\rm int}}{10^{8}\,{\rm K}}\right)^{4} } & $ T_{\rm int}\lesssim 3700\,{\rm K} $, \label{eq:small_th}
\end{subnumcases}
where Eq.~(\ref{eq:gtr_th}) is obtained from Ref.~\cite{Kouvaris:2007ay}
	and Eq.~(\ref{eq:small_th}) is the expression for $\epsilon_\gamma$ when $T_{\rm int}\lesssim 3700\,{\rm K}$.

In addition, NS heating comes from the contributions of DM annihilation and dark
kinetic heating. They are given by
\begin{equation}
\mathcal{E}_{\chi}=2m_{\chi}\Gamma_{A}=m_{\chi}C_{a}N_{\chi}^{2}f_{\chi}\label{eq:ann_heat}
\end{equation}
for annihilation and
\begin{equation}
\mathcal{K}_{\chi}=C_{c}E_{s}
\end{equation}
for dark kinetic heating. The factor $f_{\chi}$ characterizes the
energy absorption efficiency which runs from 0 to 1.
The term $E_{s}=m_{\chi}(\gamma-1)$ is 
the DM kinetic energy deposited in NS and $\gamma\simeq1.35$
\cite{Baryakhtar:2017dbj}. Therefore, we have
\begin{equation}
\epsilon_{\chi}=\frac{\mathcal{E}_{\chi}+\mathcal{K}_{\chi}}{4\pi R^{3}/3}\label{eq:e_dark}
\end{equation}
for DM emissivity.

The last quantity $c_{V}$ is the NS heat capacity of NS and is expressed as \cite{Kouvaris:2007ay}
\begin{equation}
c_{V}=\frac{T_{\rm int}}{3}\sum_{i}p_{F,i}\sqrt{m_{i}^{2}+p_{F,i}^{2}}
\end{equation}
where index $i$ runs over $n$, $p$, $e$ and the corresponding
Fermi momenta are
\begin{gather*}
p_{F,n}=0.34\,{\rm GeV}\,\left(\frac{n}{n_{0}}\right)^{1/3},\\
p_{F,p}=p_{F,e}=0.06\,{\rm GeV}\left(\frac{n}{n_{0}}\right)^{2/3}.
\end{gather*}
For calculation convenience, we have expressed all these quantities
in terms of natural unit. 


\section{SIDM implication for neutron star temperature\label{sec:numerical_result}}

Before presenting the numerical results, we briefly introduce
the setups of our calculations.
We let the time starts at $t_0=100$ years after the birth
of NS and the age of NS in our study is $t_{\rm NS}=2\times 10^9$ years.
The beginning temperature is $10^9\,{\rm K}$ for both $T_{\rm int}$ and
$T_\chi$.

In the presence of DM self-interaction,
it might come to a time $t_{s}$ that DM self-interaction cross
section reaches its geometric limit $\sigma_{\chi\chi}^{c}=\pi r_{{\rm th}}^{2}/N_{\chi}(t_s)$ where $\sigma_{\chi \chi}^c$ is some variable for our program to discriminate the DM self-capture rate attaining this limit or not. 
If the program detects that the initial $\sigma_{\chi\chi}$ input
is larger than $\sigma_{\chi\chi}^{c}$ at any time $t>t_s$, it will automatically return
$\sigma_{\chi\chi}^{c}$ to avoid overestimating the effect of DM
self-interaction.

Moreover, $\sigma_{\chi\chi}^{c}$ depends on $T_{\chi}$
due to its dependence on $r_{{\rm th}}$, Eq.~(\ref{eq:Rth}).
If DM is in thermal equilibrium with NS, then $T_\chi=T_{\rm int}$
and $T_{\rm int}$ can be used to identify $\sigma_{\chi \chi}^c$.
However, it costs some time to thermalize with NS. The thermalization timescale is given by \cite{McDermott:2011jp}\footnote{To determine $t_{\rm th}$ one needs to
	solve the DM energy loss rate: $dE/dt=-\xi n_n \sigma_{\chi n}v \delta E$. The detail
	discussion is beyond the scope of this work. It can be found in Refs.~\cite{McDermott:2011jp,Guver:2012ba,Bertoni:2013bsa} and references therein.} 
\begin{equation}
t_{{\rm th}}\approx\frac{m_{\chi}^{2}m_{n}p_{F}}{6\sqrt{2}n\sigma_{\chi n}m_{r}^{3}}\frac{1}{T_{\chi}}.\label{eq:time_th}
\end{equation}
If such timescale is longer than the age of NS, then $T_\chi$ is unable to thermalize with $T_{\rm int}$.
Thus, once our program detects $t_{\rm th}<t_{\rm NS}$ in solving the coupled differential equations Eqs.~(\ref{eq:dN}) and (\ref{eq:dT}) at some time step $t_i$, it will return $T_\chi(t_i)=T(t_i)$
and use it in the next step $t_{i+1}$ of calculation.
On the other hand, if the program finds that  $t_{\rm th}>t_{\rm NS}$ at $t_i$, it will
not only return $T_\chi(t_i)=T(t_i)$ but also identify $T_\chi(t_i)$ as the decouple temperature $T_\chi^{\rm dec}$.
No matter how time evolves, the program will recognize $T_\chi(t_{i+1})=
T_\chi(t_{i+2})=T_\chi(t_{i+\cdots})=\cdots=T_\chi^{\rm dec}$.
The DM temperature $T_\chi$ decoupled from the NS cooling curve and always stays at
$T_\chi^{\rm dec}$. 
This phenomenon is depicted in Fig.~\ref{fig:temp_int}. The explanation will be given in the
following subsection.


\subsection{Numerical results}
\begin{figure}
	\begin{centering}
		\includegraphics[width=0.45\textwidth]{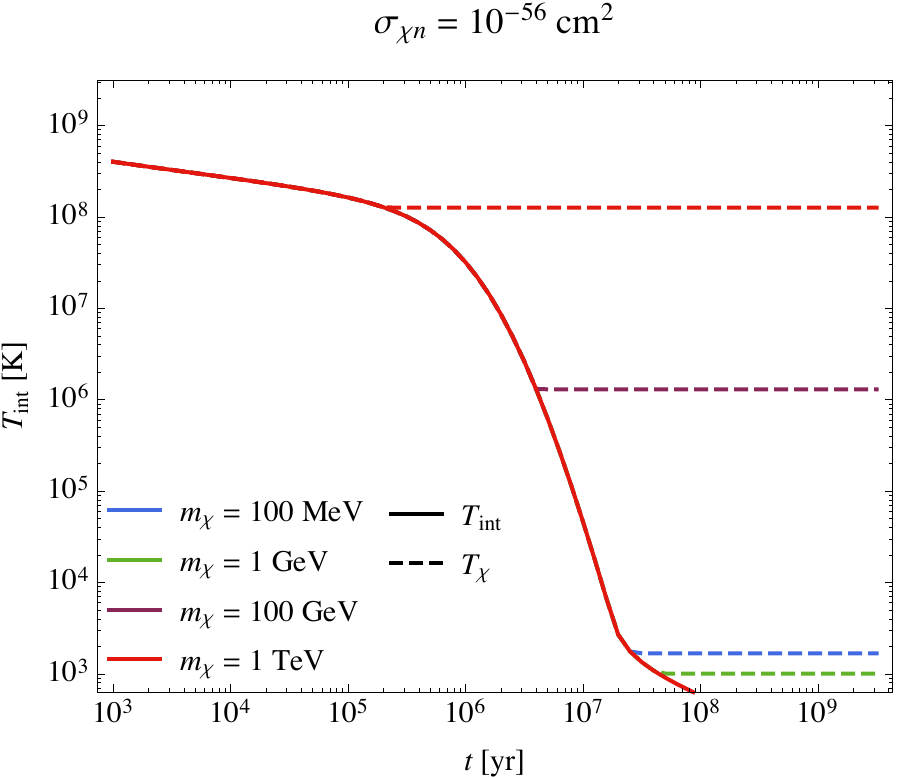}\quad
		\includegraphics[width=0.45\textwidth]{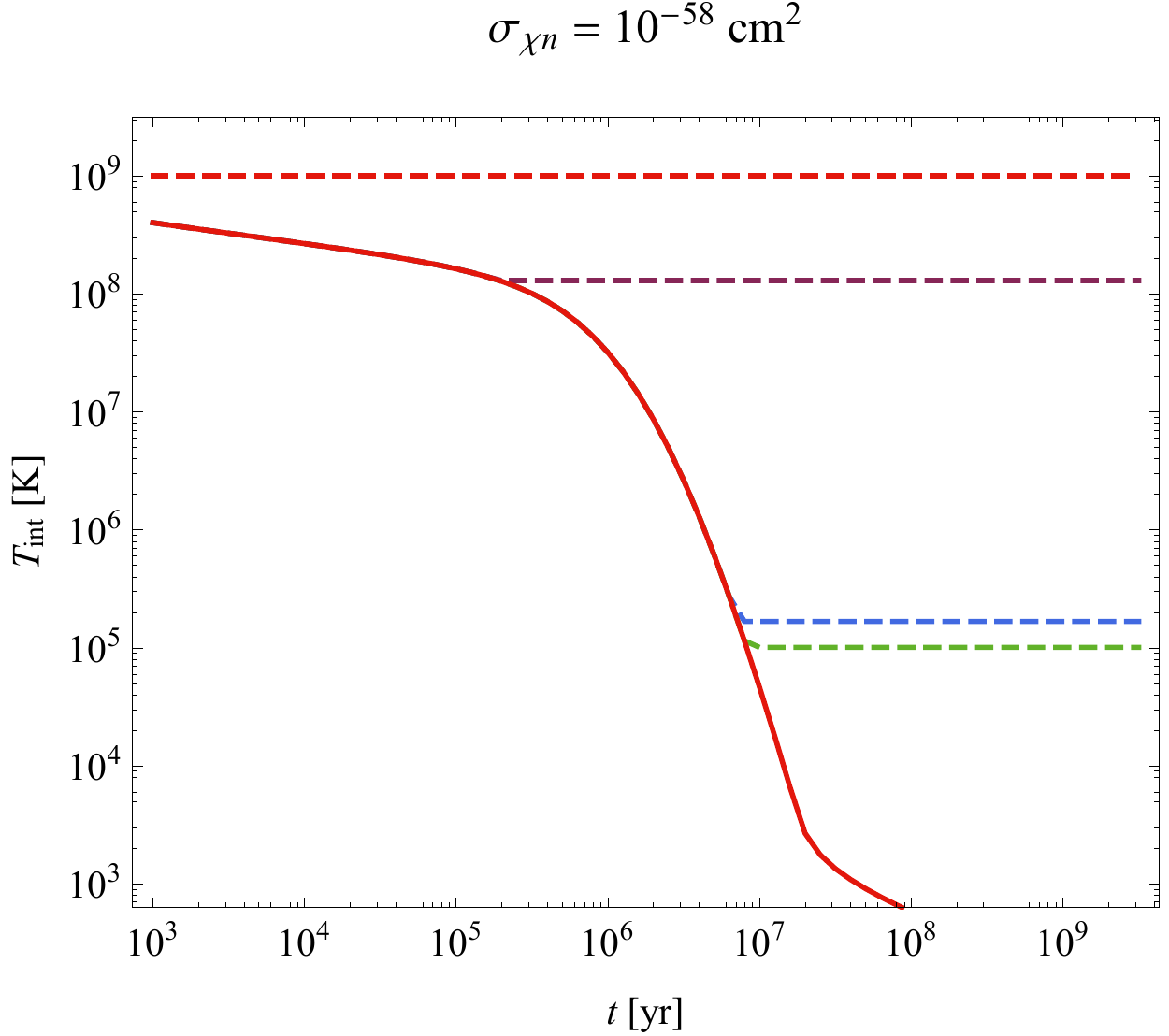}
		\par\end{centering}
	\caption{\label{fig:temp_int}The evolution of  $T_{\rm int}$
		(solid) and  $T_{\chi}$ (dashed). DM masses $m_{\chi}$
		are marked with different colors. 
		When $T_\chi$ is cold enough at some time, the corresponding $t_{\rm th}$ is longer than
		$t_{\rm NS}$. Therefore, $T_\chi$ will decouple from $T_{\rm int}$ since then (flat dashed lines).
		We use $\sigma_{\chi\chi}/m_\chi=4\,{\rm cm^2~g^{-1}}$ and $\langle\sigma v \rangle=3\times 10^{-26}\,{\rm cm^3\,s^{-1}}$ in the calculation.
		See main text for detail.}
\end{figure}
In each panel of Fig.~\ref{fig:temp_int}, the red solid line in is
	the NS cooling curve of $T_{\rm int}$, and the dashed lines are $T_\chi$. The DM masses $m_\chi$ 
	are indicated by different colors. The initial temperature is $10^9\,{\rm K}$ for both $T_{\rm int}$ and $T_\chi$. The NS cooling curve is generally insensitive to this initial condition.
	After the first few decades, the effect of initial temperature becomes negligible and this agrees with Ref.~\cite{Kouvaris:2007ay}.
	
Taking $m_\chi=1\,{\rm TeV}$ in the left panel 
	for instance, when $t\lesssim 10^5$ years, $T_\chi$ is able to thermalize with NS for a given $\sigma_{\chi p}=10^{-56}\,{\rm cm^2}$. After $t\gtrsim 10^5$ years, 
	the corresponding $t_{\rm th}$ is longer than $t_{\rm NS}$ and
	DM is unable to thermalize with NS.
	Thus $T_\chi$ decoupled from $T_{\rm int}$ since then.
	Because $t_{\rm th}$ depends on $m_\chi$ as well, $T_\chi^{\rm dec}$ is not the same for
different $m_\chi$.

For the right panel of Fig.~\ref{fig:temp_int}, it is easily
	seen that $m_\chi=1\,{\rm TeV}$ is never in thermal equilibrium with $T_{\rm int}$ from the beginning.
	This is due to $\sigma_{\chi p}$ given in this panel is too weak to have $t_{\rm th}$ smaller than $t_{\rm NS}$ even with such high initial temperature $10^9\,{\rm K}$.
Hence, the initial temperature is the maximum $T_\chi^{\rm dec}$ that DM can have.

\begin{figure}
\begin{centering}
\includegraphics[width=0.45\textwidth]{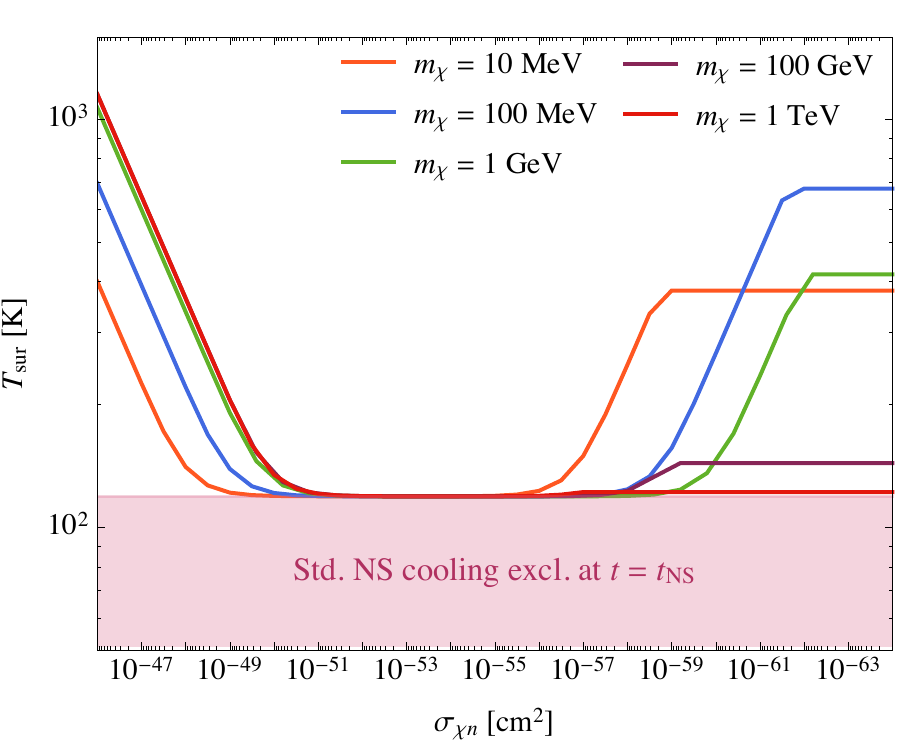}
\par\end{centering}
\caption{\label{fig:temp_sig}NS surface temperature $T_{{\rm sur}}$ versus
$\sigma_{\chi n}$ at $t=t_{{\rm NS}}$. DM masses $m_{\chi}$ are
marked with different colors
and $f_{\chi}=1$.
	We use $\sigma_{\chi\chi}/m_\chi=4\,{\rm cm^2~g^{-1}}$ and $\langle\sigma v \rangle=3\times 10^{-26}\,{\rm cm^3\,s^{-1}}$ in this calculation.
}
\end{figure}

Once NS cooling due to $\epsilon_{\nu}$ and $\epsilon_{\gamma}$ emissions
are balanced by DM heating $\epsilon_{\chi}$, the NS interior temperature $T_{\rm int}$ stops dropping.
The associated NS surface temperature $T_{\rm sur}$ for an isolated NS with age $t_{\rm NS}=2\times 10^9$
years is shown in Fig.~\ref{fig:temp_sig}.
The resulting $T_{\rm int}$ for obtaining $T_{\rm sur}$ in Fig.~\ref{fig:temp_sig}
are all smaller than 3700\,K.
Thus, there is no distinction between $T_{\rm int}$ and $T_{\rm sur}$ below this threshold.
But for convenience, we will still use $T_{\rm sur}$ in the following discussion.
See the discussion in Sec.~\ref{sec:cooling_heating}.

Without DM heating effect, a two-billion-year old isolated NS has $T_{\rm sur}\approx 120\,{\rm K}$ predicted by standard NS cooling mechanism.
For $T_{\rm sur}<120\,{\rm K}$ would be impossible.
It is indicated by the pink shaded region in Fig.~\ref{fig:temp_sig}.
When $\sigma_{\chi n}\gtrsim 10^{-50}\,{\rm cm^2}$, DM-nucleon interaction dominates the capture process and is mainly responsible for the DM heating and the deviation of $T_{\rm sur}$ from 120\,K.
While $10^{-50}\,{\rm cm^2}\lesssim \sigma_{\chi n}\lesssim 10^{-57}\,{\rm cm^2}$, neither $N_\chi$ captured through DM-nucleon interaction nor DM self-interaction can trigger enough DM heating.
The energy loss due to NS cooling, particularly from $\epsilon_{\gamma}$, overwhelms the energy injection from DM annihilation. The contribution from  $\epsilon_{\chi}$ is negligible. 

\begin{figure}
	\begin{centering}
		\includegraphics[width=0.6\textwidth]{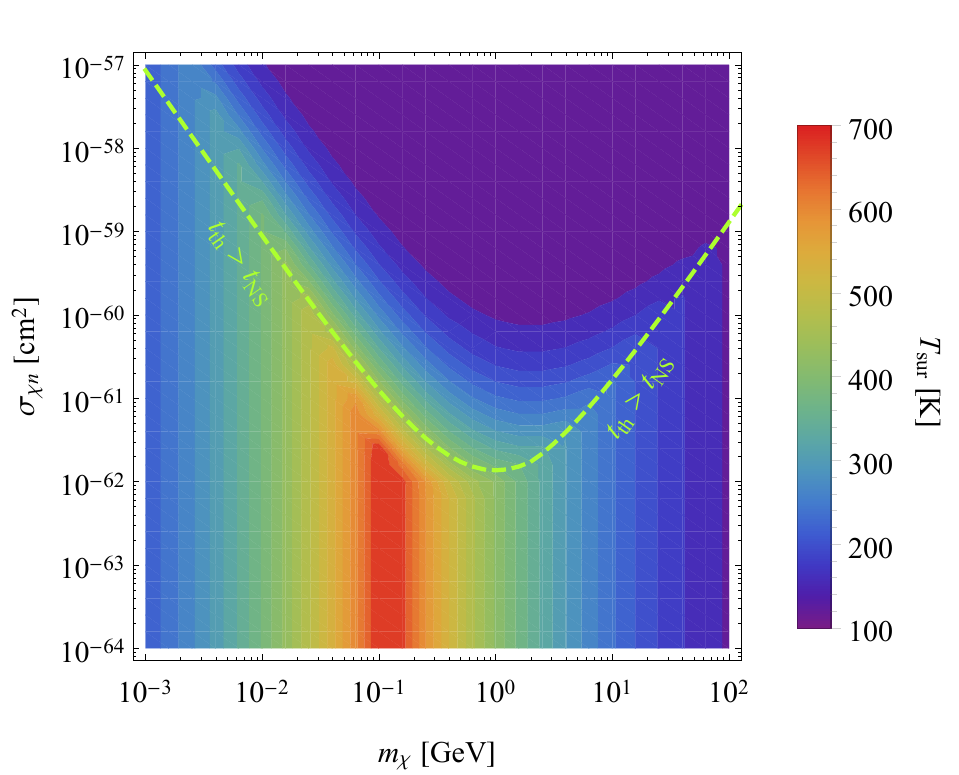}
		\par\end{centering}
	\caption{\label{fig:surf_temp}A broad scan of $T_{{\rm sur}}$ over $m_{\chi}-\sigma_{\chi n}$ plane. The corresponding DM self-interaction cross section and DM annihilation cross section are  $\sigma_{\chi\chi}/m_\chi=4\,{\rm cm^2~g^{-1}}$ and $\langle\sigma v \rangle=3\times 10^{-26}\,{\rm cm^3\,s^{-1}}$ respectively.
		See main text for detail.
	}	
\end{figure}

	When $\sigma_{\chi n}<10^{-57}\,{\rm cm^2}$, the capture process is dominated by DM self-interaction.
	For smaller $\sigma_{\chi n}$, the thermalization timescale becomes longer than $t_{\rm NS}$ hence $T_\chi$ decoupled from $T_{\rm int}$ in the earlier time with higher decouple temperature $T_\chi^{\rm dec}$. See the dashed lines in Fig.~\ref{fig:temp_int}.
	The decouple temperature $T_\chi^{\rm dec}$ decides how much $N_\chi$ will be captured inside the host star ultimately.
	Since $T_\chi^{\rm dec}$ does not change after decouple, it portrays the size of $r_{\rm th}$, Eq.~(\ref{eq:Rth}).
	The weaker $\sigma_{\chi n}$ is, the higher $T_\chi^{\rm dec}$ and the larger $r_{\rm th}$. The DM self-capture rate is stronger. More $N_\chi$ will be captured and results in greater energy injections.
	This explains the rising $T_{\rm sur}$ like a reheating  when $\sigma_{\chi n}<10^{-57}\,{\rm cm^2}$ in Fig.~\ref{fig:temp_sig}.
	One can also notice that the rising of $T_{\rm sur}$ is not unlimited.
	When $\sigma_{\chi n}$ is small enough, $T_\chi^{\rm dec}$ will not grow into larger value as its maximum is the initial temperature $10^9\,{\rm K}$.
	Such also corresponds to the maximum DM heating when the capture process is dominated by the DM self-interaction.\footnote{
		The DM self-capture rate saturates when $N_\chi \sigma_{\chi\chi}=\pi r_{\rm th}^2$.
		If $T_\chi^{\rm dec}$ stops growing larger with smaller $\sigma_{\chi n}$, then $r_{\rm th}$ ceases to increase as well as $N_\chi$ and DM heating effect.}
	The plateaus for each $m_\chi$ in Fig.~\ref{fig:temp_sig} shows the maximum $T_{\rm sur}$ caused by DM in this region.


A broad scan of $T_{{\rm sur}}$
over $m_{\chi}-\sigma_{\chi n}$ plane is displayed in Fig.~\ref{fig:surf_temp}.
The region below the green dashed line indicates the plateaus to each $m_\chi$ shown in  Fig.~\ref{fig:temp_sig}.
The NS surface temperature $T_{\rm sur}$ does not change regardless of any smaller $\sigma_{\chi n}$.
The dark purple region on the right-top is where DM heating is negligible.
The corresponding $T_{\rm sur}\approx 120\,{\rm K}$ as indicated by standard NS cooling.

\subsection{NS surface temperature as a complementary probe for DM properties}\label{sec:max_temp}
\begin{figure}
\begin{centering}
\includegraphics[width=0.6\textwidth]{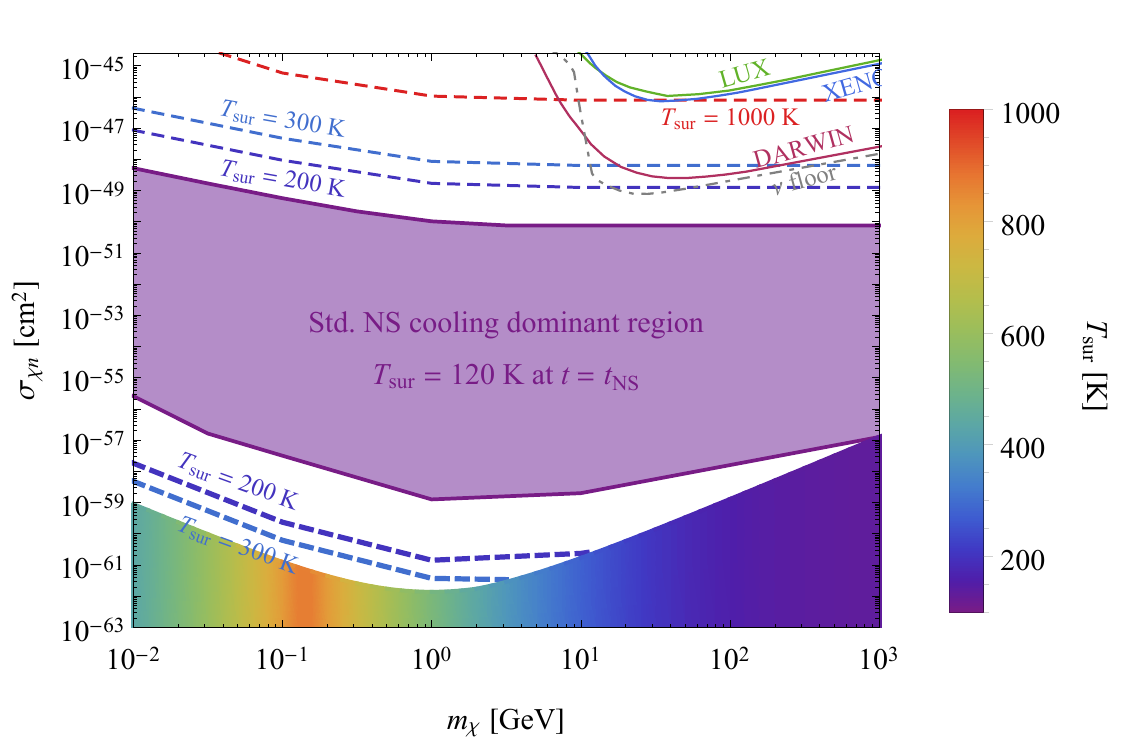}
\par\end{centering}
\caption{\label{fig:SIDM_sensitivity}
The purple shaded area
is where standard NS cooling overwhelms the DM heating.
The corresponding $T_{\rm sur}$ is about 120\,K.
DM-nucleon interaction and DM self-interaction are responsible for the heating on $T_{\rm sur}$
above and below the purple shaded region in the middle of this figure respectively.
We use $\sigma_{\chi\chi}/m_\chi=4\,{\rm cm^2~g^{-1}}$ and $\langle\sigma v \rangle=3\times 10^{-26}\,{\rm cm^3\,s^{-1}}$ in the calculation.
See main text for detail.
The constraints on $\sigma_{\chi n}$ from different DM
direct searches such as DARWIN \cite{Aalbers:2016jon}, LUX \cite{Akerib:2016vxi} and
XENON1T \cite{Aprile:2017iyp} are shown in the plot as well.}
\end{figure}

Here we display our final result in Fig.~\ref{fig:SIDM_sensitivity} as well as
the constraints on $\sigma_{\chi n}$ from different DM direct searches.
The purple shaded region is where DM heating has no contribution.
Thus, standard NS cooling mechanism predicts $T_{\rm sur}\approx 120\,{\rm K}$ for an isolated two-billion-year old NS.
Above the shaded region, the larger $\sigma_{\chi n}$, the more $N_\chi$ will be captured as well as the stronger DM heating from annihilation.
Below the shaded region, $T_{\rm sur}$ is reheated as a consequence of DM self-interaction and the color portion is the maximum $T_{\rm sur}$ can be obtained in the DM self-interaction dominant region.
This color portion in the bottom indicates the plateaus in Fig.~\ref{fig:temp_sig} and the region below the green dashed line in Fig.~\ref{fig:surf_temp}.


Interestingly, $T_{\rm sur}$ above the purple shaded region coincides with $T_{\rm sur}$ in
some parameter space below the purple shaded region.
This phenomenon is indicated in Eq.~(\ref{eq:Neq2}).
For instance, there are two lines show
$T_{\rm sur}=300\,{\rm K}$ in Fig.~\ref{fig:SIDM_sensitivity}.
It could be helpful as an extra information for determining DM properties along
with current DM direct searches.
If we observed $T_{\rm sur}=300\,{\rm K}$ for an isolated two-billion-year old NS and concur that the heating is purely from DM.
But $10\,{\rm GeV}\lesssim m_\chi \lesssim 100\,{\rm GeV}$ with $T_{\rm sur}=300\,{\rm K}$ is already disfavored by DARWIN.
Thus, DM could be either lighter than a few GeV or $\sigma_{\chi n}$ is very small. 
Future DM direct searches could reveal more constraint on sub-GeV DM, together with
the astrophysical observations on $T_{\rm sur}$, we can further unravel more information about DM.

\section{Summary\label{sec:summary}}

In this work, we found that when DM self-interaction dominates DM capture process,
$T_{\rm sur}$ increases as $\sigma_{\chi n}$ decreases.
This is contrary to DM-nucleon interaction dominant case where $T_{\rm sur}$ becomes colder as  $\sigma_{\chi n}$ diminishes.

In addition, when $ 10^{-50}\,{\rm cm^2}\lesssim \sigma_{\chi n}\lesssim 10^{-57}\,{\rm cm^2}$, the heating from DM is negligible.
The energy loss from photon emission overwhelms the energy deposition from DM annihilation.
Thus, standard NS cooling is the dominant process.
For a two-billion-year old isolated NS, standard NS cooling predicts a lower bound $T_{\rm sur}\approx 120\,{\rm K}$.
If DM properties such as $m_\chi$ and $\sigma_{\chi n}$ lie within this parameter space, they are unable to probe from the measurement of $T_{\rm sur}$.
However, the precise value for such lower bound
depends on the knowledge of NS cooling mechanism.
Once we have more constraints from the astrophysical observations on $T_{\rm sur}$ for isolated old NSs,
this lower bound could be subject to some correction.

The parameter space in the capture processes dominated by DM-nucleon interaction and by DM self-interaction can generate the same $T_{{\rm sur}}$ as shown in Fig.~\ref{fig:SIDM_sensitivity}.
Together with the current DM direct searches, it could improve our knowledge on DM properties in various ways.

In closing, the NS surface temperature $T_{\rm sur}$ induced by DM self-interaction roughly ranges from 120\,K to 700\,K.
The corresponding blackbody peak wavelength is infrared and could be detected by the
forthcoming telescopes such as JWST, TMT and E-ELT.
The corresponding observations on $T_{\rm sur}$ could act as the complementary probe to DM direct searches in the future.

\begin{acknowledgments}
C. S. Chen (TKU) and Y. H. Lin (NCKU) are supported by the Ministry
of Science and Technology, Taiwan under Grant No.~104-2112-M-032-009-MY3
and 106-2811-M-006-041 respectively.
\end{acknowledgments}

\end{document}